# Personal Information Management

William Jones[1], Jesse David Dinneen[2], Robert Capra[3], Anne R. Diekema[4], and Manuel A. Pérez-Quiñones[5]

[1]Information School, University of Washington, Seattle, WA USA

[2]School of Information Studies, McGill University, Montreal, QC Canada

[3]School of Information and Library Science, University of North Carolina, Chapel Hill, NC USA

[4]Department of Instructional Technology and Learning Sciences, Utah State University, Logan, UT USA

[5]Department of Software and Information Systems, University of North Carolina, Charlotte, NC USA

## Abstract

Personal Information Management (PIM) refers to the practice and the study of the activities a person performs in order to acquire or create, store, organize, maintain, retrieve, use, and distribute information in each of its many forms (paper and digital, in e-mails, files, Web pages, text messages, tweets, posts, etc.) as needed to meet life's many goals (everyday and long-term, work-related and not) and to fulfill life's many roles and responsibilities (as parent, spouse, friend, employee, member of community, etc.). PIM activities are an effort to establish, use, and maintain a mapping between information and need. Activities of finding (and re-finding) move from a current need toward information while activities of keeping move from encountered information toward anticipated need. Meta-level activities such as maintaining, organizing, and managing the flow of information focus on the mapping itself. Tools and techniques of PIM can promote information integration with benefits for each kind of PIM activity and across the life cycle of personal information. Understanding how best to accomplish this integration without inadvertently creating problems along the way is a key challenge of PIM.

# Introduction

*Personal Information Management (PIM) refers to the practice and the study of the activities a person performs in order to acquire or create, store, organize, maintain, retrieve, use, and distribute information in each of its many forms (paper and digital, in e-mails, files, Web pages, text messages, tweets, posts, etc.) as needed to meet life's many goals (everyday and long-term, work-related and not) and to fulfill life's many roles and responsibilities (as parent, spouse, friend, employee, member of community, etc.)*
[1]. PIM places special emphasis on the organization and maintenance of personal information collections (PICs) in which information items, such as paper documents, electronic documents, e-mail messages, digital photographs, Web page references, handwritten notes, etc., are stored for later use and repeated reuse.[2]

PIM provides a productive meeting ground for several disciplines including cognitive psychology and cognitive science; human–computer interaction, information science, and human-information behavior; and the fields of data, information, and knowledge management. In a world increasingly defined by the information we receive, send, and share, the ability to manage this information is one of life's essential skills.

One ideal of PIM is always to have the right information in the right place, in the right form and of sufficient completeness and quality to meet the current need. This ideal is far from reality for most people. In practice, people do not always find the right information in time to meet their needs. It may be that the necessary information is not found or may arrive too late to be useful. Or information may arrive too soon and then be misplaced or forgotten entirely before opportunities for its application arrive. People forget to use information even when (or sometimes because) they have taken pains to keep it stored somewhere for future use.

To better "keep found things found" [3], people may store their information in multiple locations and in multiple applications, using organizational schemes that are roughly comparable but still inconsistent (often in important ways). The result is *information fragmentation*[4] – the information people need is scattered widely, making activities of PIM more difficult:

- *Maintaining and organizing* personal information is more difficult when so many separate stores of information must be considered. Are organizations the same or at least consistent? Which version of a document is current? Is information getting backed up?

- *Managing for privacy, security and the overall flow of information* (incoming and outgoing) is more difficult when a person's information is so widely scattered. What have people shared and with whom? [5] Is there an "Achilles heel" in a person's data protection efforts?

- Information fragmentation makes more difficult even the basic, every-minute activities of *keeping* and *finding.* Where/how to keep an important file, reference or reminder when there are so many alternatives to choose from? And where later should people look?

Information fragmentation is manifest in the many sources of information a person must consult to make even simple decisions. The decision to accept a dinner invitation, for example, may depend upon the information to be found in digital calendars, paper calendars (e.g., a shared household calendar), Web pages, and e-mail exchanges. PIM activities in general take longer and are more prone to error when information is scattered.

While new applications, tools, and devices often increase information fragmentation, this does not have to be the case. Some tools and techniques of PIM promote information integration, which can simplify PIM activities and have benefits across the lifecycle of personal information. Tools can also simplify the sharing and exchange of information across devices and people, and can help users to see unified views of their information spaces.

In the following sections, we provide:

- ➢ a brief history of PIM,
- ➢ an analysis of PIM including the several senses in which information can be "personal" and major kinds of PIM activity,
- ➢ a selective review of PIM research especially as this relates to major kinds of PIM activity, and
- ➢ concluding notes on some of the significant challenges and opportunities for the PIM field of study in the years ahead.

# A Brief History of PIM

PIM is a new field with ancient roots. When the oral rather than the written word dominated, human memory was the primary means for information preservation [6]. As information was increasingly rendered in paper form, tools were developed over time to meet the growing challenges of management. For example, the vertical filing cabinet, now such a standard feature of home and workplace offices, was first commercially available in 1893 [7].

The modern dialog on PIM is often traced to the publication of "As We May Think" by Vannevar Bush at the close of World War II [8]. Bush expressed a hope that technology might be used to extend our collective ability to handle information and to break down barriers impeding the productive exchange of information.

The 1940s also saw the development of Shannon's theory of communication which laid the groundwork for a quantitative assessment of information [9]. Key to this theory is the notion that the information content of a message can be measured for its capacity to reduce uncertainty.

To make information exclusively about the reduction of uncertainty can be seen as overly restrictive [10]. As Buckland notes, "sometimes information increases uncertainty" [11, p. 351]. Nevertheless, even as we search for complementary, qualitative characterizations of information (e.g. [12]) a larger point in Shannon's work endures: the value of information is not absolute, but relative to a context that includes the intentions of the sender, method of delivery, and the current state of a recipient's knowledge. This holds true even when sender and recipient are the same person albeit separated in time as, for example, when people place "appointments" in their calendars a week from today as a reminder to do something (e.g. "book plain tickets").

With the increasing availability of computers in the 1950s came an interest in the computer as a source of metaphors and a test bed for efforts to understand the human ability to process information and to solve problems. Newell and Simon pioneered the computer's use as a tool to model human thought [13], [14]. They produced "The Logic Theorist," generally thought to be the first running artificial intelligence (AI) program. The computer of the 1950s was also an inspiration for the development of an *information processing approach* to human behavior and performance [15].

After the 1950s research showed that the computer, as a symbol processor, could "think" (to varying degrees of fidelity) like people do, the 1960s saw an increasing interest in the use of the computer to help people to think better and to process information more effectively. Working with Andries van Dam and others, Ted Nelson, who coined the word "hypertext" [16], developed one of the first hypertext systems, The Hypertext Editing System, in 1968 [17]. That same year, Douglas Engelbart also completed work on a hypertext system called NLS (oN-Line System) [18]. Engelbart advanced the notion that the computer could be used to augment the human intellect [19], [20]. As heralded by the publication of Ulric Neisser's book *Cognitive Psychology*[21], the 1960s also saw the emergence of cognitive psychology as a discipline that focused primarily on a better understanding of the human ability to think, learn, and remember.

The computer as aid to the individual, rather than remote number cruncher in a refrigerated room, gained further validity from work in the late 1970s and through the 1980s to produce personal computers of increasing power and portability[1]. These trends continue. Computational power roughly equivalent to that of a desktop computer of a decade ago can now be found in devices that fit into the palm of a hand[2].

The phrase "Personal Information Management" was itself apparently first used in the 1980s [26] in the midst of general excitement over the potential of the personal computer to greatly enhance the human ability to process and manage information. The 1980s also saw the advent of so-called "PIM tools" that provided limited support for the management of such things as appointments and scheduling, to-do lists, phone numbers, and addresses. And a community dedicated to the study and improvement of human–computer interaction also emerged in the 1980s [27], [28].

As befits the "information" focus of PIM, PIM-relevant research of the 1980s and 1990s extended beyond the study of a particular device or application towards larger ecosystems of information management to include, for example, the organization of the physical office [29] and the management of paperwork [30]. Malone [31] characterized personal organization strategies as 'neat' or 'messy' and described 'filing' and 'piling' approaches to the organization of information. Other studies showed that people vary their methods for keeping information according to anticipated uses of that information in the future [32]. Studies explored the practical implications that human memory research might carry in the design of, for example, personal filing systems [33]–[35] or information retrieval systems [36]. Studies demonstrated a preference for navigation (browsing, "location-based finding) in the return to personal files [37], a preference that endures today notwithstanding significant improvements in search support [38]–[41] and an increasing use of search as the preferred method of return to e-mails[42].

PIM, as a contemporary field of inquiry with a self-identified community of researchers, traces its origins to a Special Interest Group (SIG) session on PIM at the CHI 2004 conference [43] and to a special National Science Foundation (NSF)-sponsored workshop held in Seattle in 2005 [44].[3]

# An Analysis of PIM

A deeper understanding of what PIM is begins with definitions of associated concepts and the description of a conceptual framework in which to interrelate key kinds of PIM activity.

---

[1] For more on the history of the personal computer see[22]–[24]. For more on the history of the personal computer see [22]–[24].

[2] See Jones [25] , chapter 4, for discussion on the present and future of information "always at hand" via devices.

[3] Since 2005, PIM workshops have been held roughly every 18 to 24 months (http://pimworkshop.org/).

## The Information Item and Its Form

Information sent and received takes many *information forms* in accordance with a growing list of communication modes, supporting tools and people's customs, habits and expectations. People still send paper-based letters, birthday cards and thank you notes. But ever more so, people communicate using digital forms of information including e-mails, documents shared (as attachments or via a service such as Dropbox), and "tweets", text messages, blog posts, Facebook updates and personal Web pages.

Across forms, it is useful to speak of an *information item* as a packaging of information. An item encapsulates information in a persistent form that can be managed. An information item can be created, stored, moved, given a name and other properties, copied, distributed, and deleted.

The concepts of information item and information form are helpful abstractions for considering aspects of PIM. For example, an e-mail message, as an item, packages together information such as a sender, receiver, subject, body text, and date sent. All e-mails, as befits their form, share in common characteristics such as the expectation of a timely delivery (i.e., in seconds not days) and a timely response (i.e., within a day or two, not a month later). An e-mail message can be *transformed,* i.e., it can be copied to create a new information item of a different form. For example, a user may print the e-mail and view it in a paper form. Each form is associated with distinct set of tools, techniques, habits and expectations for interacting with and managing the information.

## Personal Information

There are several senses in which information can be personal. Each represents a different relationship between the information and the person: Information can be owned by, about, directed toward, sent by, experienced by, or relevant to "me." (See Table 1). This definition and the several relationships captured in Table 1 form an inclusive definition of Personal Information. Prior work in PIM has at times limited the definition of PIM by not including some of these relationships.

*Table 1. Information can be personal in any of several relationships to a person.*

|    | **Relation to "me"**             | **Examples**                                                                                                                                                 | **Issues**                                                                                                                                                                                                                  |
|----|----------------------------------|--------------------------------------------------------------------------------------------------------------------------------------------------------------|-----------------------------------------------------------------------------------------------------------------------------------------------------------------------------------------------------------------------------|
| P1 | Controlled by/owned by me        | E-mail messages in our e-mail accounts; paper documents in a home office; computer files on a hard drive or in a Web cloud service.                          | Security against break-ins, virus protection, backups to prevent data loss etc. What to share and with whom? These questions apply both for paper filing cabinets and for folders in a service like Dropbox.               |
| P2 | About me                         | Credit history, medical records, web search history, records of library books checked out.                                                                   | Who sees what when and under what circumstances? How is information corrected or updated? Does it ever go away?                                                                                                             |
| P3 | Directed toward me               | E-mails, phone calls, drop-ins, TV ads, web ads, pop-ups. Also, invitations to join a share of a cloud service. Facebook posts. Tweets "@" you.              | How to stay focused long enough to complete a task? "When I'm not interrupted by others, I interrupt myself! I worked non-stop today, but what did I accomplish?"                                                           |
| P4 | Sent/posted/ shared by me        | Sent e-mail, personal web sites, published reports and articles, folders shared via a cloud service like Dropbox.                                            | Who sees what when? Did the message get through? Am I making the "right" impression? Am I sharing the right information for our teamwork?                                                                                   |
| P5 | For things experienced by me     | Web history, photos (taken by others as well as by me), hand-written notes, full-motion videos from head-mounted cameras[45].                                | How to get back to information again later? How to pick up where I left off (i.e., with a task, a book or episodes of a drama on Netflix)? How much history do I really need?                                               |
| P6 | Potentially relevant/ useful to me | Somewhere "out there" is the perfect vacation, house, job, lifelong mate. If only I could find the right information!                                        | If only I knew (had some idea of) what I don't know. But also... how to filter out information we don't wish to see? (How to do likewise for our children?)                                                                 |

Clearly the senses in which information can be personal are not mutually exclusive. In many cases, the same information may be personal in several senses. For example, a folder of photos taken last Thanksgiving is information about the friends and family at the dinner – and about "me" (P2). The photos are owned by "me" (P1) and these represent events experienced by "me" (P5). The owner can also elect to share the photos with others via a service like Dropbox (P4).

## A Personal Space of Information

Each person has a single unique personal space of information (PSI) defined by the union of personal information in each of its six senses as depicted in Table 1. A PSI affects the way its owner views and interacts with the rest of the world. A PSI also affects the way its owner is seen, categorized, and treated by others.

At its center, a person's PSI includes all the information items that are, at least nominally, under that person's control. At its periphery, the PSI includes information that the person might like to know about and control but that is under the control of others. Included is information about the person that others keep. Also included is information in public spaces, such as a local library or the Web, which might be relevant to the person.

## Personal Information Collections

Personal information collections, referred to as PICs or simply collections, are personally managed subsets of a PSI. PICs are "islands" in a PSI where people have made some conscious effort to control both the information that goes in and how this information is organized. PICs can vary greatly with respect to the number, form, and content coherence of their items. Examples of a PIC include:

- The papers in an office and their organization, including the layout of piles on a desktop and the folders inside filing cabinets.
- A collection of projects each represented by a folder stored in a cloud storage service and accessed from different devices.
- A collection of information items related to a specific project that are initially "dumped" into a folder on a person's notebook computer and then organized over time as the project takes shape.
- A reference collection of articles in digital format, organized for repeated use across projects.
- Digital songs managed through a laptop computer or smartphone.
- A collection of media (TV shows, movies, music) stored in a streaming service under a personal account, organized by some preference and including information such as ratings, watched/unwatched, and other notes.

A PIC includes not only a set of information items but also their organizing representations including spatial layout, containing folders, properties, and tags.

## Activities of PIM

*PIM activities* are an effort *to establish, use, and maintain a mapping between information and need*, whether actual, perceived, or anticipated. Figure 1 depicts a mapping between information and need as well as the activities of PIM that work with this mapping.

Information can arrive as aural comments from a friend, a billboard seen on the way to work, letters in surface mail or any number of digital items including e-documents, e-mail messages, Web pages, text messages and tweets.

Needs can arise from internal or external sources. A person may recall, for example, that she needs to make plane reservations for an upcoming trip. The need may arise from the question posed by a colleague in the hallway or a manager's request during a meeting. External needs are also be evoked by an information item, such as an e-mail message or a Web-based form.

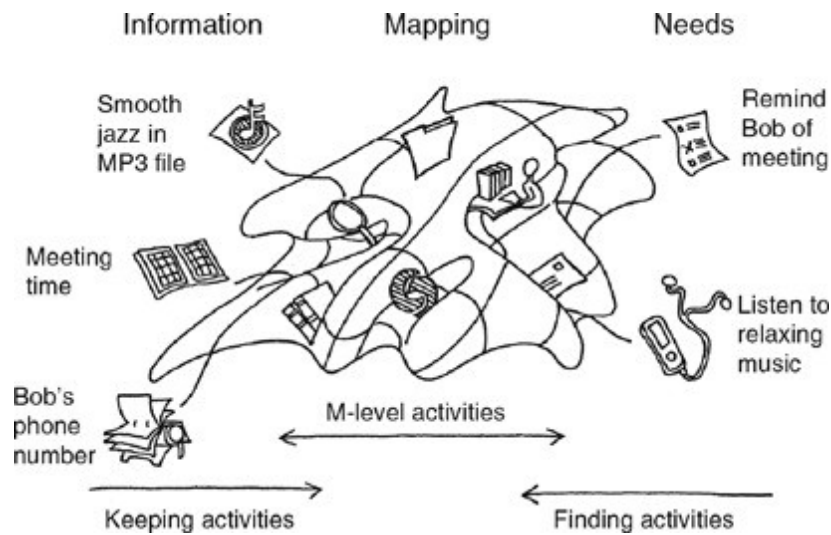

Fig. 1 PIM activities viewed as an effort to establish, use, and maintain a mapping between needs and information. Keeping and finding activities are interrupt driven (as prompted by incoming information or a need). M-level (meta-level) activities are broader in focus and more strategic in nature[4].

Connecting between need and information is a mapping. Only small portions of the mapping have an observable external representation. Much of the mapping has only a hypothesized existence in the memories of an individual. Large portions of the mapping are potential and not realized in any form, external or internal. A sort function or a search facility, for example, has the potential to guide from a need to the desired information.

However, parts of the mapping can be observed and manipulated. The folders of a filing system (whether for paper or digital information), the layout of a desktop (physical or virtual), and the choice of names, keywords, and other properties for information items all form parts of an observable fabric helping to knit need to information.

Activities of PIM can be grouped and interrelated with reference to Figure 1 as follows:

- **Keeping activities** map from encountered information to need (immediate or anticipated). This grouping includes decisions concerning whether attend to information in the first place and, then, whether to make any effort to keep the information. Some information – the time on a clock or a sports score, for example -- is "consumed" immediately with no need to keep the information for later access. Much of the information people encounter is simply ignored. For the subset of encountered information to be kept for later use, questions arise regarding how the information should be kept. Should information items be piled (where?), filed (which folder?), tagged (with which tags?), or committed to memory?
- **Finding/re-finding activities** map from need to information. This grouping includes explicit search queries as posted to a Web-based search service or to a computer desktop-based search facility. The grouping also includes various activities of sorting, navigation, and "nosing around" that people use to get back to information for reuse.
- **Meta-level activities** focus on the mapping itself as a way of connecting together information and need. Meta-level activities include efforts to maintain (through backups, periodic cleanups, updates,

---
4

and corrections) and organize (via schemes of piling, filing, or tagging) information; manage privacy, security, and the flow of information (e.g., through subscriptions, friendships, policies of disclosure, firewalls, virus protection); measure and evaluate (how are supporting tools and strategies working?); and make sense of and use of personal information ("What is the information telling me? What should I do?").

# PIM Research

PIM research can be organized according to PIM activity as follows:

## Finding/Re-finding: From Need to Information

The focus of a finding/re-finding activity can be on information in a public space such as a physical library or the Web, or focus can be on the information a person owns in a private space such as an office filing cabinet, the hard drive of a personal computer, or a cloud storage service.

A large body of information seeking and information retrieval research applies especially to finding public information [46]–[50]. There is a strong personal component even in efforts to find new information, never before experienced, from a public store such as the Web. For example, efforts to find information may be directed by a personally created outline or a to-do list. In addition, information inside a person's PSI can be used to support a more targeted, personalized search of the Web [51].

The search for information is often a sequence of interactions rather than a single transaction. Bates [52] describes a "berry picking" model of searching in which needed information is gathered in bits and pieces through a series of interactions and during this time, the user's expression of need, as reflected in the current query, evolves. Teevan et al. [53] note that users often favor a stepwise approach even in cases where the user might have sufficient knowledge to access the information more quickly and more directly via a well-formed query. The stepwise approach may preserve a greater sense of control and context over the search process and may also reduce the cognitive burden associated with query formulation. For frequent information needs, users may well-trod paths of access, whereas for less common needs, search may be more likely to play a role [54].

People may find (rather than re-find) information even when this information is ostensibly under their control. For example, items may be 'pushed' into the PSI (e.g., via the inbox, automated downloads, Web cookies, the installation of new software, etc.), and a person may have no memory for or awareness of these items. If these items are ever retrieved, it is through an act of finding, not re-finding. Memories for a previous encounter with an information item may also fade so that its retrieval is more properly regarded as an act of finding rather than re-finding [55]. While searching for information, one might come across information that, though not relevant to the current search, is relevant for another project. Or one may find information that is incorrectly filed or is in a "download" folder awaiting more proper organization (e.g., as part of "spring cleaning").

Lansdale [26] characterized the retrieval of information as a two-step process involving interplay between recall and recognition. The steps of recall and recognition can iterate to progressively narrow the search for the desired information—as happens, for example, when people move through a folder hierarchy to a desired file or e-mail message or navigate through a Web site to a desired page.

But a successful outcome in a re-finding effort depends upon completion of another step preceding recall: a person must remember to look. A person may know exactly where an item is and still forget to look for the item. It is also useful to consider a final "repeat?" step. Failure to collect a complete set of information

(e.g., all potential conflicting commitments before accepting a proposed meeting time) can mean failure for the entire finding episode.

Re-finding, then, is a four-step process with a possibility to fail at each step:

**Step 1. Remember (to look)**. Many opportunities to re-find and reuse information are missed for the simple reason that people forget to look. This failure occurs across information forms. In a study by Whittaker and Sidner [56] for example, participants reported that they forgot to look in "to do" folders containing actionable e-mail messages. Because of mistrust in their ability to remember to look, people then elected to leave actionable e-mail messages in an already overloaded inbox.

Web information is also forgotten. In one study of Web use, participants complained that they encountered Web bookmarks, in the course of a "spring cleaning" for example, that would have been very useful for a project whose time had now passed [57]. In a study where participants were cued to return to a Web page for which they had a Web bookmark, the bookmark was used on less than 50% of the trials [58]. Marshall and Bly [59] report a similar failure to look for paper information (newspaper clippings).

Remembering to look depends, in part, upon the effective use of *attentional spaces* [3, Ch. 4] such as the surface of a physical desktop, a computer desktop, or the visible region of an inbox's display.

**Steps 2 and 3. Recall and recognize**. Recall and recognition can be considered together as two sides of a dialog between people and their PSIs. A person types a search word (recall) and then scans through a list of results (recognition). A person's recall is often partial. For example, a person may have some sense that the e-mail message was sent by a person whose name starts with an "S" ("Sally?"). As prompted by this partial recall, she might then sort e-mail messages by sender and then scan through the "s"-section of this sorted list.

Though any act of retrieval involves a combination of recall and recognition, their relative importance can vary with method. Recall is typically more important in search i.e., a person initially recalls the keywords to be used in a search query (though, of course, the person might have recognized these words from a web page previously viewed and simply copied them into the query). Recognition is typically more important in what is alternately referred to as location-based finding, orienteering, browsing or, simply, *navigation* [37], [53], [60].

Many PIM studies have determined that users prefer to return to information, most notably the information kept in personal digital files, by navigating rather than searching [37]–[39], [41], [53], [60]–[62]. This may be surprising; the use of navigation often depends on remembering or recognizing a specific path to the folder in which the item was stored, while search is more flexible insofar as users can specify in their query any attribute they happen to remember about the information item. Fertig et al. argued back in 1996 [63] that an observed preference for navigation over search was due to the limitations in the search technologies of at the time, and that improvement in search would inevitably lead to a shift from navigation to search as the preferred method of return. Commercial PIM search engines have indeed improved dramatically. However, the combined results of a large-scale study and a longitudinal study [39] indicated that the availability of new and improved search facilities (such as Mac Spotlight) did not result in more search than was observed for older and slower search facilities (such as Sherlock). Search was apparently used only as last resort on those occasions when participants could not remember the location of the files they sought. This finding gains further support from a recent log study indicating that participants used search in only 4.2% of their retrievals [64].

One explanation for a preference for navigation over search in the return to personal information derives from a persistent, general finding in cognitive psychology literature that people are better able to recognize than to recall information previously encountered. (see [21]. As noted above, in the interplay between recognition and recall, recognition is more dominant in navigation and recall is more dominant in search.

"Navigation" to digital files and other information is, of course, done in a computing environment that is much less rich in sensory experience than the physical environments people navigate through on a daily basis. But perhaps digital and physical navigation engage, at some level, some of the same processes? In support of this intuition is a recent, innovative fMRI study [65] demonstrating that, when people use navigation as a means to retrieve their own personal files, they are using some of the same brain structures as when navigating in the physical world (i.e., the posterior limbic regions). In contrast, search activates Broca's area, commonly observed in linguistic processing.

Another recent study asked people either to navigate or search to a target file. As people did so, a list of words was read to them in the background. People were later better able to remember these words, presumably a linguistic task, when navigating rather than searching[66].

The argument for the observed preference for navigation to search in the retrieval of personal information would then appear to be very strong indeed. Navigation places higher reliance on recognition, which people are usually better at than recall. Navigation to digital information engages some of the same neural structures as are engaged in physical navigation, which is a highly practiced skill in people. Furthermore, navigation places fewer demands than search on linguistic processing and the regions of the brain involved in linguistic processing thus freeing this capacity for use elsewhere in a person's computing interactions.

But the discussion continues. Just as navigation vs. search, and the use of supporting tools, appear to engage different facilities and different areas of the human brain, there is reason to suppose that other activities, too, as done through the computer and computer-based tools will vary qualitatively in the nature of processing demands placed upon the user. Studies done in the 1980's, for example, suggested a significant shift in processing demands with the advent of support for full-screen, WYSIWYG text editing [67]–[69].

A more recent study suggests that the now-standard support for fast, indexed "desktop" search may be having some impact, not on the preferred method for return to personal digital files, which remains navigation, but on another form of personal information: E-mail [42]. Consistent with this is the finding that people are less inclined to invest in the organization of e-mails into folders electing instead, for example, to leave information in a single larger folder or even the inbox[70].

How, then, to account for this apparent disparity — that even as preference may be shifting towards search as a primary method for return to e-mails (i.e., those that aren't readily available in the current inbox display) — people persist in navigating to digital files? And why this despite the fact that essentially the same fast, indexed search support can be applied to both e-mails and files?

We consider one last explanation that relates to the well-established *primacy* effect of psychology research[21]: Under a *first-impressions* hypothesis [42], people's first choice of return aligns with their initial experience with the encountered information. Given adequate search support, people might naturally prefer to retrieve an e-mail via a search on sender or subject since this information is often an initial focus of attention (and is used as a basis for deciding whether to attend to the e-mail). Likewise, since people must initially place a file into a folder (unless they accept a "My Documents" default), navigation to the folder is a first choice of return to the file. Given better, more universal support for tagging, for example, people might be more likely to prefer search (on tags) as a preferred method of return to documents and other files. (But see the section below on

Keeping for its discussion of folders vs. tags). Conversely, were people first to create and exchange e-mails "in place", in the context of a project (folder) they are working on, they might prefer navigation as a method of return. (See, for example, work on the Project Planner/Planz representing one attempt to situate e-mail by project[71], [72]).

Habits change slowly and desktop search support continues to improve. Desktop search is increasingly integrative and ever closer to an ideal where anything that can be remembered about an information item or the circumstances surrounding encounters with it (e.g., time of last use or nearby events) can be used to help find this item [35], [73], [74].

Search vs. navigation? Since many circumstances of real-world information retrieval are likely to involve a little of both [53], a challenge in our tool-building efforts towards better PIM is to support artful ways of combining and seamlessly switching between these two general methods of information retrieval. Towards this goal, is recent work that explores ways to integrate search and navigation by highlighting certain navigational pathways based upon the results of a background search [75].

**4. Repeat?** In many instances, information need is not for a single information item but, rather, for a set of items whose members may be scattered in different forms within different organizations. If the likelihood of successful retrieval of each item is strictly independent from the others, then the chances of successful retrieval of all relevant items goes down as their number increases. So even if the likelihood of success for each item of four items is high, retrieval of all four items is presumably much lower.

However, in situations of *output interference*, items retrieved first interfere with the retrieval of later items so that chances of successful retrieval of the whole set are worse than predicted by a strict independence of individual retrievals [76]. The effect can be explained with reference to interference paradigms of human memory retrieval such as the basic *spreading activation model*[77]. As a useful simplification, think of items in a simple set as primarily indexed and accessed through a single node such as "People in my book club". In this case, during a retrieval attempt, each item in the set competes with other items for a limited amount of activation. Moreover, the successful retrieval of initial items strengthens their connections to the access node at the expense of later items not yet retrieved e.g., as "Susan" is retrieved, the corresponding link to "People in my book club" is strengthened to the expense of, say, the link to "Jonathan" (not yet retrieved).

The chances of successfully retrieving all members of a set can also be much better than predicted by independence. Certainly, retrieval is better if all items are in the same larger unit—a folder or a pile, for example. Retrieval may also be better if items of a set have an internal organization or are interconnected to one another so that the retrieval of one item actually facilitates the retrieval of other items.[78]. By one explanation, activation sent to one item, can be "shared" with other items through these interconnections [79].

## Keeping: From Information to Need

Many events of daily life are roughly the converse of finding events: People encounter information and try to determine what, if anything, they should do with this information, i.e., people must match the information to current or anticipated needs. Decisions and actions relating to encountered information are collectively referred to as *keeping* activities.

The ability to handle effectively information that is encountered by happenstance may be essential to a person's ability to discover new material and make new connections [80]. People also keep information that they have actively sought, but do not have time to process currently. A search on the Web, for example, often produces much more information than can be consumed in the current session. Both the decision to keep this information for later use and the steps to do so are keeping activities.[57]

Keeping activities are also triggered when people are interrupted in the midst of a current task and look for ways of preserving current state so that work can be quickly resumed later.[81] For example, people keep appointments by entering reminders into a calendar, and keep good ideas or "things to pick up at the grocery store" by writing down a few cryptic lines on a loose piece of paper.

People keep not only to have the information later but also build in a reminder look for it. A failure to remember to use information later is one kind of prospective memory failure.[82]–[84] People may, for example, self-e-mail a Web page reference in addition to or instead of making a bookmark because the e-mail message with the reference appears in the inbox where it is more likely to be noticed and used. [57]

Research relating to keeping points to several conclusions: 1) keeping is difficult and error-prone; 2) keeping "right" has gotten harder as the diversity of information forms and supporting tools has increased; and 3) some costs of keeping "wrong" in the digital realm have gone away due to advanced search capabilities, but challenges remain.

**Keeping is difficult and error-prone**. The keeping decision is multifaceted. Is the information useful? If so, do special steps need to be taken to keep it for later use? How should the information be kept? Where? On what device? In what form should the information be kept? The keeping decision can be characterized as a signal detection task [4] subject to errors of two kinds: 1. An incorrect rejection ("miss") when information is ignored that, as time tells, should have been kept (e.g., those paper receipts that are needed now that the tax return is being audited). 2. A false positive when information kept as useful ("signal") turns out not never to be used later. The physical and digital clutter in our lives is often evidence for false positive decisions.

Filing information items—whether paper documents, e-documents, or e-mail messages—correctly is a cognitively difficult and error-prone activity [26], [31], [56]. Difficulty arises in part because the definition or purpose of a folder or tag is often unclear from its label (e.g., "stuff") and then may change in significant ways over time [56], [85]. Filing and tagging difficulties increase if people do not recall the folders or tags they have created previously and then create new folders or tags rather than reusing those already created [56].

But placing (or leaving) information items in piles, as an alternative to filing, has its own problems. In Malone's study[31], participants indicated that they had increasing difficulty keeping track of the contents of different piles as the number of piles increased. Experiments by Jones and Dumais [86] suggest that the ability to track information by location alone is quite limited. Moreover, the extent to which piles are supported for different forms of information is variable, limited, and poorly understood [87].

Kwasnik [32] identified a large number of dimensions that might potentially influence the placement and organization of paper-based mail and documents in an office. Overall, a document's classification was heavily influenced by the document's intended (anticipated) use—a finding subsequently reproduced by Barreau [88]. Similarly, Jones et al. [57] observed that the choice of method for keeping Web information for later use was influenced by a range of considerations or functions. Marshall and Bly [59] also note that the reasons for keeping information vary and are not necessarily task-related or even consciously purposeful.

**Keeping "right" is harder when information is more fragmented**. An act of keeping might be likened to throwing a ball into the air toward a point where we expect to be at some future point in time and space. [2] The ongoing proliferation of information forms and supporting tools and gadgets makes keeping activities even harder. The information people need may be at home when they are at work or vice versa. It may be on the wrong device. Information may be "here" but locked away in an application or in the wrong format so that the hassles associated with its extraction outweigh the benefits of its use. Users employ a variety of methods to ensure that information is in the "right" place when it is needed and methods may require additional, manual work to transfer and synchronize information across devices and locations. Personal- and employer-imposed information boundaries may further fragment the information owned by or under the control of the user (P1 – see Table 1) with significant impact on activities to maintain and organize and also to manage the flow of information [89]. (Considerations for these and other *meta-level activities* are further described later in this section).

People must contend with the organization of e-documents, e-mail messages, Web pages (or references to these), and possibly also a number of additional forms of digital information including phone messages, digitized photographs, music, and videos—each with its own special-purpose tool support[90]. The number of keeping considerations further increases if a person has different e-mail accounts, uses separate computers for home and work, or uses any of a number of special-purpose PIM tools.

**Keep less, keep nothing, keep more, keep everything or keep "smarter"?** With respect to some forms of information, people may elect to adopt a "keep less" (or even a "keep nothing") approach. For example, people can rely on "do nothing" methods of return to Web pages—such as searching, auto-completion, inspection of browsing history or navigating from another Web site—that require no explicit keeping activity [58].

On the other hand, experiencing difficulty in re-finding can motivate people towards a "keep more" approach. After a failure to re-find a useful Web site, for example, teachers were reported to have shifted towards a more aggressive strategy of bookmarking "anything" that looked at all useful [91]. But conscious keeping is costly in time, distraction and the potential clutter of new items especially when these end up to be "false positives" i.e., information judged to be useful but never in fact used [4].

Digital memory research (e.g., [92]) explores an automated "keep everything" or "total capture" approach manifest in one extreme by a full-motion, line-of-sight video recording of daily experiences, second by second [45]. But the utility of such an approach can be questioned on grounds both practical (e.g., how to selectively access just a portion of the continuous stream?) and psychological (the recording alone does not compare to human memory, and its ability to evoke a person's actual memory for events may decline with time) [93, p. Chapter 6], [94]. Automated keeping also points to a dilemma identified by Lansdale[26]: without some engagement in the keeping process people may be much more likely to forget about the information later.

Alternatives to "keep everything," "keep nothing," and "keep automatically" approaches are ones that help people to "keep smarter," i.e., to make better decisions concerning future uses of current information [4]. If people have prepared a clear plan, for example, they are often more effective at keeping relevant information (including a recognition of its relevance) even when the plan and its goal are not the current focus of attention [95].

## Folders or Tags?

The demise of folders and the file system is a recurrent theme in discussions on the future of information management (e.g., [96], [97], see also [98, Ch. 6 & 7]). There are good reasons for such predictions. A strict folder hierarchy does not readily allow for the flexible classification of information even though, in a person's mind, an information item might fit in several different categories[99]. For example, pictures from a conference in Copenhagen can be stored under "Pictures," "Trips," "Conferences," or "Copenhagen." Why not then, instead of placement in a single folder, allow people to tag these items i.e., with a tag for each of the categories above?

Tagging models of information storage would seem to align more closely with the way people think and might wish to represent their information. Towards the exploration of this possibility, a number of tag-related prototypes for PIM have been developed over the years (e.g., [100]–[103]). A tagging approach has also been embraced in commercial systems, most notably Gmail (as "labels").

The verdict? Success of tags so far is at best mixed. In one study, Bergman et al. [104] found that users when provided with (and informed of) both options to use folders and tags, prefered folders to tags (for both their files and e-mails), and even when using tags they typically refrained from adding more than a single tag per information item (also a finding in [105]). Even Gmail, though fundamentailly acrhitected to support tagging, has added support for "folders"[5].

Why isn't the case for tagging more compelling (so far)? Support for the tagging (annotating, labeling) of information items remains basic and fragmented. Many design questions remain. One study, for example, found that users performing tagging had more persistent memories about their information, but that nuanced interaction design was necessary to alleviate the cognitive demand entailed by creating and recalling tags [106].

But more fundamental limitations may also apply. Civan et al. [105] through an engagement of participants in critical, comparative observation of both tagging and the use of folders (through use of Grmail and Hotmail for the organization of e-mails) were able to elicit some limitations of tagging not previously discussed openly in comparisons between tagging and the use of folders. For example, once a user decides to use multiple tags it becomes important to continue doing so. A tag such as "Copenhagen" not consistently used becomes much less useful later as a retrieval cue. Conversely, critical discussion can reveal advantages in the use of folders that are not always apparent such as, for example, that a folder structure can serve as an informal task decomposition and a simple kind of reminder for task completion[107].

Just as the answer for the finding (retrieval) of information may be neither navigation nor search but tools that support an artful combination of both, so to the answer towards better keeping may be neither purely tagging nor use of folders but a tool that artfully combines the best of both.

## The Meta-level: Mapping between Need and Information

Meta-level activity operates broadly upon collections of information (PICs) within the PSI, the configuration and effective use of supporting information tools, and on the mapping that connects need to information. The daily actions of keeping and finding are largely reactive (i.e., faced with a need, find the information to meet this need; faced with information, identify if and what need it might meet). In contrast, meta-level activity is more proactive. Thus, meta-level activity focuses not on the need or information immediately at hand but on a larger consideration of information collections, supporting tools, and the overall mapping. Alternatively, meta-level activity can be seen to move to a more strategic level in a people's efforts to manage their information

One factorization of meta-level activity makes the simplification that the PSI is one big store to be maintained and organized, and for which input and output should be managed. People might also benefit from an ability to take measurements in support of various evaluations of the store (e.g., does it have the right information in the right forms, sufficiently current and complete?). This approach leads to the following overlapping kinds of meta-level activity: *Maintain and organize; manage privacy, security and the flow of information* (incoming and outgoing); and *measure and evaluate.* Implicit in each of these activity types is an effort to *make sense of and use* the available information. This section takes a close look at each of these meta-level activity types.

---

[5] http://googlesystem.blogspot.com/2009/02/gmail-adds-folders-by-improving-label.html.

But first, note the following:

➢ Actual PIM behavior we would wish to observe and analyze is likely a combination of meta-level activity, keeping, and finding. The sharing and synchronization of information in a cloud service such as Dropbox, for example, has elements of *maintenance and organization* (i.e., information is automatically backed up) and *managing for privacy, security and information flow* (can people trust that their information is secure and safe from snooping? With whom should people share what?). The meta-activity types we consider here might be thought of as vectors – independent but not necessarily orthogonal from one another – forming a basis for the description of a "point" of actual PIM behavior (in a large space of possibilities).

➢ A consideration of meta-level activity makes clear the point that information is not an end in itself but rather a means to an end. Going back to the definition of PIM at this article's outset, PIM is about managing information to realize life's goals and roles. Further, information provides people with a way of managing their limited resources. Here we can consider "META" as an acronym for money, energy, time and attention. Each of these resources must be managed if people are to be effective in the realization of their goals and the fulfillment of their roles and responsibilities. But people don't "touch" (manage) these resources directly. People manage via their information. Financial statements tell people whether their expenses are under control and whether they are on track for retirement. Their calendars help them to manage their time. The use of open windows and the tabs of a Web browser can be considered as a way of managing attention.

➢ Further and especially as people's information is increasingly in digital form, people don't "touch" (manage) their information directly either but rather via their information tools. (Even for information in paper form, people use tools, of course, including staplers, filing cabinets and their own hands). An investment to select, configure and learn to use information tools is a critical component of each kind of PIM activity. This holds especially for meta-level activity.

➢ Finally, as another sense of "meta", meta-level activity is the "after" activity. Actions at this level such as "spring cleaning" or "clean-up", notwithstanding their overall importance, are seldom urgent and so easily postponed and avoided.

Patterns of postponement and avoidance are in evidence as we further explore meta-level activity in the remainder of this section.

## Maintaining and organizing.

Differences between people are especially apparent in their approaches to the maintenance and organization of information. Malone [31] distinguished between "neat" and "messy" organizations of paper documents. "Messy" people had more piles in their offices and appeared to invest less effort than "neat" people in filing information. Comparable differences have been observed in the ways people approach e-mail [56], [108]–[110], e-documents [58], [62], and Web bookmarks [62], [111].

Across information forms, differences in approaches to organization correlate with differences in keeping strategy. For example, people who have a more elaborate folder organization—whether for paper documents, e-documents, e-mail messages, or Web bookmarks—tend to file sooner and more often [62]. However, people are often selective in their maintenance of different organizations. Boardman and Sasse [62], for example, classified 14 of 31 participants in their study as "pro-organizing" with respect to e-mail and e-documents but not with respect to bookmarks; 7 of 31 participants only took the trouble to organize their e-documents. Diekema and Olsen [91] found that Spring cleaning of information by teachers mostly pertained to physical information rather than digital information. Organization behaviors may also vary based on context. In a study by Capra et al.[112] university employees reported having more folders for their work e-mail accounts than for their personal accounts.

Several studies have now looked at how the same person manages across different forms of information [62], [107], [113].The following composite emerges:

- ➢ People do not generally take time out of a busy day to assess their organizations or their PIM practice in general.

- ➢ People complain about the need to maintain so many separate organizations of information and people complain about the fragmentation of information that results. People struggle to organize their information so that they can keep their focus of attention and avoid "getting lost" [114]–[116].

- ➢ Even within the same personal information collection (PIC), competing organizational schemes may suffer an uneasy coexistence with each other. People may apply one scheme on one day and another scheme the day after.

- ➢ Several participants in one study reported making a special effort to consolidate information in their PICs [57], for example, by saving Web page references and e-mail messages into a file folder organization or by sending e-documents and Web page references in e-mail messages.

- ➢ People don't, in general, have reliable, sustainable (i.e., automatic) plans for backing up their information [117], [118].

Even if digital forms of information can be integrated, people must still contend with paper forms of information. Paper documents and books remain an important part of the average person's PSI. [85]. The integration of paper and digital forms of information can be troublesome. In a study of elementary and secondary teachers, for example, Diekema and Olsen [91] found that the teachers' dualistic system of digital and physical information was especially challenging; while some teachers tried to standardize their organization schemes across material type, some teachers resorted to digitizing all materials. Others printed out their digital materials so they could be filed with the rest of the paper files.

## Managing privacy, security and the flow of information.

We continue with the useful simplification that a person's PSI is one large store. If maintenance and organization activities are concerned with the store itself and its contents, the activities to manage privacy, security and, more generally, the flow of information are concerned with the input and output to the store. What do people let in? What do people let out?

Letting the wrong things (information, data) into a PSI or letting the wrong tools access or modify a PSI can, at minimum, be a major hassle (e.g., a need for "disinfection"). At worst, the computer may be hijacked to nefarious ends and all its data corrupted. Indeed, problems of malware ("viruses", "Trojans", "worms") are endemic. By some estimates 30% or more of the computers in the United States are infected [119]. Letting the wrong information out to the wrong people can also be a costly source of trouble – e.g., if credit cards are compromised or worse a person's identify is stolen.

And yet, the exchange of information, incoming and outgoing, and increasingly in digital forms, is an essential part of living in the modern world. To order goods and services on-line people must be prepared to "let out" their credit card information. To try out a potentially useful, new information tool, people may need to "let in" a download that could potential make unwelcome changes to the web browser or the desktop. Providing for adequate control over the information, coming into and out of a PSI, is a major challenge. Even more challenging is the user interface to make clear the implications for various choices in privacy control and in clicking the "Accept" button for use of services such as Facebook [120]–[128]

Also of relevance are the daily challenges associated with sharing information with others in various social situations ranging from the home, to school, to work, to play and "at large" (e.g., via open-ended discussion boards on the Web). These challenges are further discussed in the section on "GIM and the social fabric of PIM".

### Measuring and evaluating.

Choices are made in support of all the activities described so far. Schemes of organization are selected; strategies, policies and procedures are adopted; supporting tools are put in place. People then need to ask, at least occasionally, "Is it (the resulting mapping between information and need) working? Is it helping me to make the best use of my limited resources (money, energy, time, attention) towards meeting my goals and fulfilling my roles and responsibilities? Can it work even better? If so, what should change?"

These questions depend both upon the measurements that can be made and also on the evaluations people must make in cases where measurements (and the underlying objectives they reflect) are in competition with one another.

Actions of measuring and evaluating are often high-level and qualitative. For example, as people are driving home from work, much later than they'd planned for and with no time, nor energy, to exercise at the gym, they may reflect on a day of non-stop meetings and interruptions as they ask themselves "What did I _really_ accomplish?"

But increasingly evaluations can be based upon numerical data. Questions such a "where does the money go?" or "where did the time go today?" needn't be only rhetorical. Increasingly detailed, consolidated data is becoming available to answer these and other questions concerning not only how people manage their resources but also how effectively they're living the lives they wish to live. Detailed financial data can tell people "where" the money is going. Calendars consolidated with activity logs can tell people "where" the time went and much more besides. On which devices? Which applications? Relating to which people or which projects? Involving physical activity as well? Or, conversely, the intake of calories (e.g., a business lunch)? How did physical measures of heart rate, blood pressure, blood sugar levels, etc. vary during these activities?

People's every actions can be tracked and a digital trace formed. The nearly constant use of computational devices (even as people sleep) creates opportunities to capture additional data about people and their environment for purposes of correlation and, potentially, deeper understanding. Associated issues of privacy are enormous and beyond the scope of this article to address (but see [129]–[131]).

On a more positive side, how can measurements be used to a person's advantage?

The increasing capture (by intention or incidentally) and availability of data in digital form to measure everything from heart beats to gaze duration to affective response give rise to notions of the *quantified self* and *personal informatics* [132]–[137]. Definitions of personal informatics or PI vary. Some would tightly connect PI to enabling devices and applications for tracking (e.g.,[138]). But self-efforts to track personal activity – the better to understand and make adaptive changes in behavior -- are not new[6] . Rapp and Cena provide a more flexible definition: "**Personal Informatics** *(PI), also known as Quantified Self (QS), is a school of thought which aims to use technology for acquiring and collecting data on different aspects of the daily lives of people."* [140, p. 613].

PI applications are often affiliated with health information and many times have a dedicated hardware device, such as the FitBit[7]. But some examples also include systems that collect personal information over time from a variety of sources to provide upon request more integrative summaries of the many scattered events happening that relate to a person in one way or another. A popular example of this second variety is Mint.com which collects information from all of one's financial accounts (checking, saving, credit cards, etc.) and summarizes this data into an integrated dashboard.

Personal informatics is properly regarded as a subset of PIM where special focus is given to activities of measuring and evaluation and also to making sense of and using the data collected. For example, Li et al. [141] decompose a personal informatics system into five stages: preparation, collection, integration, reflection, and action. The first three stages fit well under the more general activity of measuring and evaluating while the last two stages fit well under the more general activity of making sense of and using information. Not surprisingly, users of PI systems also face challenges to manage for privacy and information flow and to maintain and organize the large amounts of data collected. According to Li et al. [142], users face information fragmentation problems comparable to those already discussed in the introduction to this article. More recent work in personal informatics seems to acknowledge a need to think more broadly and to place efforts at "self-tracking" in the larger context of an overall PIM practice [134], [143], [144].

## Making sense of and using information.

Making sense of information represents another set of meta-level activities operating on personal information and the mapping between information and need. People must often assemble and analyze a larger collection of information in order to decide what to do next. For example, among the choices available in new automobiles (new laptops, new employees, etc.), which will best meet a person's needs? Which treatments to select in the aftermath to surgery for cancer? Does radiation therapy "make sense"? Making sense of and using information means more than just retrieval and comprehension of the information "at hand". A person may look up a budget number and understand that it reads "42" but may conclude that the number doesn't "make sense" (and is possibly a pop culture reference[8]) if all other budget numbers are 9 digit numbers beginning with "11-".

Making sense of information is "meta" not only for its broader perspective but also because it permeates most PIM activity even when the primary purpose may be ostensibly something else. For example, people organize information in folders ostensibly to insure its subsequent retrieval but then also as a way of categorizing and so making sense of the information. In the Jones et al. study [107], folder hierarchies developed for a project often resembled a project plan or partial problem decomposition in which subfolders stood for project-related goals and also for the tasks and subprojects associated with the achievement of these goals. Similarly some teachers organize their files by utilizing the structure as established in their curriculum standards [91].

Barsalou [145], [146] has long argued that many of a person's internal categories arise to accomplish goals. His research demonstrates an ability of people to group together seemingly dissimilar items according to their applicability to a common goal. For example, weight-watchers might form a category "foods to eat on a diet." Rice cakes, carrot sticks, and sugar-free soda are all members of the category, even though they differ considerably in other ways.

---

[6] See, for example, the description of Benjamin Franklin's efforts at self-observation in ([139, p. Chapter 8])
[7] http://www.fitbit.com/.
[8] https://en.wikipedia.org/wiki/42_(number).

Folders and tags (and piles, properties/value combinations, views, etc.) can form an important part of external representations which, in turn, can complement and combine with internal representations ("in our heads") to form an integrated cognitive system [147], [148]. Finding the right external representation can help in *sense-making* [149], [150] i.e., in efforts to understand the information. For example, the right diagram can allow one to make inferences more quickly [151]. The way information is externally represented can produce huge differences in a person's ability to use this information in short-duration, problem-solving exercises.[152] Different kinds of representations, like matrices and hierarchies, are useful in different types of problems [153]–[155].

But the effects of an external representation on longer-term efforts to plan and complete a multi-week or multi-month project are less well understood. Mumford et al. [156] note that, more generally, the study of planning has proceeded in "fits and starts" over the past 50 years and remains underdeveloped. Of special relevance to the study of PIM is the possibility that better support of planning might also, as a by-product, lead to better organizations of the information needed to complete a project.

## GIM and the social fabric of PIM

*Group Information Management (GIM)* as a natural extension to PIM has been written about elsewhere [157], [158]. The study of GIM, in turn, has clear relevance to the study of Computer Supported Collaborative Work (CSCW) [159].

In legal contexts it may be useful to treat a corporation as an entity – a "person" – in its own right. A corporation or other organization may have policies, procedures, habits, history and routines that further give it a collective identity as an entity in its own right.

But in many situations, there really is no group entity operating on its own behalf to manage information towards realization of goals and fulfillment of roles. Rather, there is a collection of individuals. Members of a group may hold in common (more or less) certain goals (whether implicit or explicitly expressed in a charter or mission statement). Members may apportion group roles (e.g., treasurer, secretary, president, etc.) but each member operates as an individual and according to personal interest. In keeping with this article's topic, therefore, focus is more appropriately on what might be called the *social fabric of PIM* [160, Ch. 10]. Needs of the individual must be met in any group situation as, for example, when engineers need their own private space for working drafts they are not yet ready to share with a larger team [161]. On the other hand, group and social considerations frequently enter into a person's PIM strategy as, for example, when one member of a household "keeps" medical test results by passing these on another member of the household who has agreed to maintain and organize medical information for all household members. (And perhaps there is a reciprocal arrangement for financial information).

Although a collaborative "divide and conquer" approach (where responsibilities are apportioned among the members of a group) can reduce the burdens of PIM, the opposite is frequently the case when people seek to share information collections. People may vary greatly in their approaches especially to keeping and maintaining and organizing information [162]. Working together through these issues often entails a delicate negotiation process [89], [163], as categorization affects each member of the shared space, and the tools used to perform such activities provide no preexisting conventions for tasks like organizing a project's files and folders [164]. Some users may customize shared information spaces to meet their own needs, but at the cost of decreasing the intelligibility of that space for others [165]. These challenges later affect re-finding information in shared spaces, where attempts to retrieve information from a shared collection are more likely to fail than when retrieving from a PIC that is not shared [166]

As further studies aim to better understand the challenges of sharing and managing information in group contexts, improved software may help to alleviate some of these challenges and facilitate the relevant activities [5]. Prototypes of such software have attempted to support group information management, for example by analyzing group activities [167]; a 'silver bullet' solution has yet to be found, however. When faced with the limitations of current software, users often prefer more traditional, ad hoc methods of sharing information, such as the use of e-mail attachments[168] to more sophisticated methods supported by tools more explicitly designed to support collaborative work such as the use of Web services that support shared folders[166], [169] In some cases people will circumvent group-centered approaches even when they are institutionalized [170]. Therefore the need for understanding and improving collaborative information tasks is clearly great, and work remains to be done.

## The interplay between PIM activities

We noted in the previous section that the senses in which information can be personal (i.e., owned by, about, directed towards, sent/shared by, experienced by or relevant to "me") don't provide for groupings of information that are cleanly separated from one another. Rather these are better seen as different perspectives on one large PSI.

Likewise, activities of PIM can be seen as different vectors of approach, we might say, to a person's PSI and to the creation, use and maintenance of a mapping (between need and information) for this space. Any actual action of PIM is likely to be a mixture of each. The interplay between activities of finding and keeping was noted in the previous section. An act of keeping, for example, often engenders acts of finding (e.g., the effort to find a folder where the encountered information should be kept). Likewise, an act of finding frequently engenders ancillary acts of keeping. In the effort to retrieve a file from a personal store, for example, the person may move ("re-keep") other files after noticing these are in the wrong folder. And certainly, newly found information beyond simple look-ups (e.g., the weather report or a sports score) must often be kept pending the proper time and place for more careful study.

Interplay is in further evidence as the meta-level of PIM activity considered. Any actual activity of PIM observed or reported is likely to resist clean categorization as solely one or another kind of activity. People may elect to place project-related files, web references and even e-mails into a single folder, named for the project [71], [171], [172] and may elect to keep this folder under a larger folder synced with a service such as Dropbox. The folder may then be shared with others also involved in completion of the project. Keeping activity? Maintaining and organizing? Making sense of and using information? Managing for the flow of information (and privacy/security)? Yes: all four. And since a cloud service such as Dropbox maintains a log of activity in the folder, we can add Measurement and evaluation.

# Conclusion: Challenges and Opportunities

To summarize, the practice of PIM involves finding, keeping, and several kinds of meta-level activity in an effort to establish, maintain, and use a mapping between information and need. Challenges arise with respect to each kind of activity:

**Finding is a multistep process**. First, people must *remember* to look. An item is retrieved through variations and combinations of searching and navigation. Navigation and searching each involve an iterative interplay between basic actions of *recall* and *recognition*. Finally, in many situations of information need, people must *repeat* the finding activity several times in order to "re-collect" a complete set of information items in order then to make sense of and use the information at hand.

**Keeping is multifaceted**. Certainly keeping, like finding, can involve several steps. But the essential challenge of keeping stems from the multifaceted nature of the decisions about information needs. Is the information useful? Do special actions need to be taken to keep it for later use? Where? When? In what form? On what device? With no crystal ball to see into the future, answering these questions is difficult and error-prone. Caution is warranted with regards to tool support to "keep everything" or "keep automatically". The approach can lead to the collection and storage of information that overwhelms even as it falls far short of providing a true "external memory".

**Meta-level activities are important but easily overlooked**. Meta-level activities are critical to a successful PIM practice, but rarely urgent. There are few events in a typical day that direct a person's attention to meta-level activity involving considerations of maintenance and organization, managing privacy, security and the flow of information, measuring and evaluating, or making sense and using information in a larger collection. As a result, meta-level activities can easily become after-thoughts.

Researchers also face several challenges in their efforts to develop PIM as a field of study:

**Integration to counter fragmentation**. In their daily practices of PIM, people must overcome problems both of information overload and information fragmentation. Problems of information fragmentation are often introduced by the very computer-based tools that are meant to help with PIM. The study of PIM itself is often fragmented in similar ways. Many excellent studies focus, for example, on e-mail, other studies on the Web, or the management of special forms of digital information (such as digitally encoded songs or photographs). There is a need for integrative approaches both in the study of PIM and in the design and evaluation of supporting tools. The "I" in "PIM" should be for information in all its many forms, paper-based and digital, and for a larger perspective that considers the life cycle and impact of information as it is communicated from one person to the next.

**Towards more practical methodologies with more practical implications.** PIM requires the study of people, with a diversity of backgrounds and needs, over time as they work in many different situations, with different forms of information and different tools of information management. This scope of PIM inquiry brings a need for practical, cost-effective methodologies that can scale. Further, there is a need not only for *descriptive* studies aimed at better understanding how people currently practice PIM (e.g., [173]) but also for *prescriptive* studies aimed both at evaluation [174] and also towards the recommendation of proposed solutions in the form of new, improved tools, techniques and strategies of PIM.

A kind of "checkbox" methodology – call it *"six by six (6x6)"* --can be devised from the framework described in this article (first described in [139] and updated in [1]) with its provision for 6 senses in which information can be personal (1. controlled by/owned by, 2. about, 3. directed toward, 4. sent/posted/shared by, 5. for things experienced by, 6. potentially relevant/useful to "me") and its 6-part factorization of PIM activity (1. keeping, 2. finding, 3. maintaining and organizing, 4. managing privacy, security and the flow of information, 5. measuring and evaluating, 6. making sense of and using information).

For any tool, technique or larger strategy of PIM, real or proposed, it is important to consider likely impact on personal information in each of its senses. For example, a "cool new tool" (desktop or web-based) that promises to deliver information potentially "relevant to me" (P6, the "6$^{th}$ sense" in which information is personal) may do so only at the cost of a distracting increase in the information "directed to me" (P3) and by keeping too much personal information "about me" (P2) in a place not under the person's control. Similarly, a tool, technique or strategy of PIM can be considered for its impact on PIM activity. Tools of personal informatics and digital tracking, for example, may do a great deal to improve "measuring" but if such improvements are accompanied by extra hassles in maintaining and organizing or in managing privacy, security and the flow information, then the tradeoffs may not be worth it.

This 6x6 methodology is not the only approach towards a more thoughtful, systematic design of PIM tools. Bergman, for example, reports good success in the application of a *user-subjective approach* in PIM system design [175]. The user-subjective approach advances three design principles. In brief, these are that design should allow that 1. all project-related items no matter their form (or format) are to be organized together (the *subjective project classification principle);* 2. the importance of information (to the user) should determine its visual salience and accessibility (the *subjective importance principle*); and 3. information should be retrieved and used by the user in the same context as it was previously used in (the *subjective context principle).* The approach may suggest design principles that serve not only in evaluating and improving existing systems, but also in creating new implementations. For example, according to the *demotion principle*, information items of lower subjective importance should be demoted (i.e. making them less visible) so as not to distract the user, but be kept within their original context just in case they are needed. The principle has been applied in the creation of several interesting prototypes[176]–[178].

In a similar vein is research by E. Jones [179] identifying five factors that determine whether a promising new system (tool, technique, strategy) of PIM will be successfully adopted: 1. *Visibility* – more generally do users notice the system and remember to use it?; 2. *Integration* – does the system connect and work well with what the user already has in place? 3. *Co-adoption* – is it necessary for others to use the system as well? (Lack of co-adoption is a major source for failure in CSCW applications [180]); 4. *Scalability* – does the system still work over time and especially as the amount of information continues to increase? 5. *Return on investment (ROI) –* after the hopeful enthusiasm with which a system is initially embraced has waned, the system has to prove its value and that the costs of its use are more than compensated for by benefits. The trouble with the adoption of PIM systems is twofold: 1. some may be discarded prematurely before the daily habit of their use has been established (e.g., for lack of visibility or integration). 2. the problems with other systems (e.g., lack of scalability or return on investment) are apparent only much later after a system has been deeply embedded in a person's overall practice of PIM making its removal a costly hassle. Consider, for example, a "perfect" folder organization that turns out to be too much trouble to maintain but can't be undone without many hours of re-organization.

More recently the notion has been advanced that tools, techniques and strategies of PIM might be assessed for their ability to increase or reduce the "clerical tax" people pay as they manage their information [93, p. Chapter 9]. What is the impact on the time spent in clerical, mechanical actions of PIM (think filing paper documents in a vertical filing cabinet) and how much time remains for the more fun, creative aspects of working with information? The utility of such an approach will depend upon the development of more detailed methods for reliably, effectively "assessing" clerical tax.

Finally, are efforts to engage the user in active discussions (possibly mediated by the Web) directed towards the evaluation and even the design of PIM systems? Use of participant observation is well-established in the fields such as sociology and anthropology (see, for example, [181]). But here we might talk instead about the *observant participation* i.e., explicitly tasking a participant to reflect upon a PIM system, the good and the bad, either in a single session or over an extended period of time. The use of diary methods of data gathering is well-established (for example, [81]) though motivating people to consistently record their thoughts can prove challenging and reminders can prove intrusive or, worse, impact the data gathered [182].

Two recent efforts of *observant participation* are worthy of further mention. Civan et al. [105] engaged users in a within-subjects comparison of tagging (labeling) vs. the use of folders as a means of organizing e-mails. Participants experienced each condition (via actual web-based versions of Hotmail and Gmail as these were offered in 2007/2008) in an ordering counter-balanced across participants. After each condition (each lasting for five days), participants were interviewed using an open-ended set of questions to assess their experiences of and reflections on the condition just experienced (i.e., use of folders vs. tags for organization of information). The study provided a very useful set of considerations and some unexpected drawbacks in the use of tags (e.g., with the freedom to tag items in several ways comes the potential tedium of then consistently using multiple tags for _every_ new item). A second method – following a Delphi approach [183] -- shows great promise for better understanding various current practices of PIM and assessing which to "recommend" and "advise against" [184].

Observant participation relates to but is distinct from approaches in *participatory design* [185], [186] wherein that attempt to "actively involve all stakeholders (e.g. employees, partners, customers, citizens, end users) in the design process to help ensure the result meets their needs and is usable"[9]. Methods of observant participation such as those described above are best seen as exploratory heuristics to gather pros and cons and considerations in the use of one tool, technique, system or strategy of PIM vs. another. The participant observer approach to PIM can be traced at least back to Malone's [31] work (with only 10 participants). In his "methodological note", Malone writes "*Sometimes… carefully controlled studies or more extensive naturalistic observations are suggested by the insights obtained from exploratory observation, and these are certainly worth performing. In other cases, the needs for designing systems (or time and budget constraints) do not justify other studies*."(p. 101).

**How to protect the privacy and security of personal information**? The more complete the digital records of personal information, the more completely a person's identity can be stolen. New PIM tools—especially those aimed at information capture—must be accompanied by new levels of information security and privacy [127].

Other questions arise for PIM-related tools and technologies:

**How to keep e-mail from being spoiled by its own success**? E-mail messages are easy and free to send—good for us; good for spammers too. E-mail is used for tracking tasks, storing documents, and saving contact information. It was designed for none of these. People may feel like they are "living" in their inbox[187]. But are they doing the things they really want to be doing? Why should all incoming correspondence go into one undifferentiated inbox? E-mail filters to sort have their own problems (e.g., that people forget to look in the sorting folders later). Research efforts have explored ways to better situate e-mails and other correspondence in larger contexts involving tasks, projects and "places" of collaboration (e.g., [188]–[190]). People are also making ad hoc use of alternatives to e-mail such as Facebook groups [191].

**How can search get more personal**? Search has application to PIM both as an interaction and as an underlying technology. Search as an interaction supports finding and re-finding and also other PIM activities such keeping, maintaining, and organizing. Search as technology, including the analysis and indexing of information, can be seen more broadly as a way of "mining" a PSI towards a wide range of personalized support [51], [93, p. Chapter 7], [192, p. Chapter 11], [193].

**How can information convergence bring information integration**? Information tools increasingly let people work with their information anytime and anywhere. This has increased the amount and diversity of information collected. Thoughts of laptop, smartphones or tablets as a point of convergence for information and practices of PIM are inspired by the ongoing rapid pace of technical developments in hardware capacity, miniaturization, and integration.

---

[9] http://en.wikipedia.org/wiki/Participatory_design.

Thoughts of the Web as a point of convergence are also inspired by hardware advances supporting, for example, increases in bandwidth (especially along the "last mile" to private homes) and storage capacity. But equally, the Web is a point of convergence for its basic ability to connect, nearly instantaneously, person to person and person to information, no matter where these are in the real world, no matter what the physical distances. But convergence does not equal integration. Web-based and portable computing are still in their infancy, and new opportunities for convergence and integration are still being created (e.g., watches, wearables). The field of PIM has a special opportunity to guide development in these new areas.

As researchers study and strive to improve PIM, it is important never to lose sight of the "personal" in PIM. People manage information neither for its own sake nor for the joy of managing. PIM is a means to an end. A person manages information in many forms, ranging from appointments of a digital calendar to paper checks of a check book, in order to manage life's resources—money, energy, attention and, especially, time. A person manages information as a means toward the fulfillment of a life's important goals and precious roles.